\documentclass[english]{article}
\usepackage[T1]{fontenc}
\usepackage[latin9]{inputenc}
\usepackage{textcomp}
\usepackage{amssymb}
\usepackage{esint}

\makeatletter
\newcommand{\lyxaddress}[1]{
\par {\raggedright #1
\vspace{1.4em}
\noindent\par}
}

\makeatother

\usepackage{babel}

\begin{document}

\title{Entropy Corrections for a Charged Black Hole of String Theory}

\author{Alexis Larrañaga}

\maketitle

\lyxaddress{National University of Colombia, National Astronomical Observatory.\\
Bogota, Colombia.}

\lyxaddress{ealarranaga@unal.edu.co}

\begin{abstract}
We study the entropy of the Gibbons-Maeda-Garfinkle-Horowitz-Strominger
(GMGHS) charged black hole, originated from the effective action that
emerges in the low-energy of string theory, beyond semiclassical approximations.
Applying the properties of exact differentials for three variables
to the first law thermodynamics we derive the quantum corrections
to the entropy of the black hole. The leading (logarithmic) and non
leading corrections to the area law are obtained.
\end{abstract}



\section{Introduction}

When studying black hole evaporation by Hawking radiation using the
quantum tunneling approach , a semiclassical treatment is used to
study changes in thermodynamical quantities. The quantum corrections
to the Hawking temperature and the Bekenstein- Hawking area law have
been studied for the Schwarzschild, Kerr and Kerr-Newman black holes
\cite{otros,mann} as well as BTZ black holes \cite{akbar}. 

It has been realized that the low-energy effective field theory describing
string theory contains black hole solutions which can have properties
which are qualitatively different from those that appear in ordinary
Einstein gravity. Here we will analyse the quantum corrections to
the entropy of the Gibbons-Maeda-Garfinkle-Horowitz-Strominger (GMGHS)
charged black hole \cite{gmghs}, which is an exact classical solution
of the low-energy effective heterotic string theory with a finite
amount of charge. To obtain the quantum corrections we use the criterion
for exactness of differential of black hole entropy from the first
law of thermodynamics for two parameters. We find that the leading
correction term is logarithmic, while the other terms involve ascending
powers of inverse of the area. 

In the quantum tunneling approach, when a particle with positive energy
crosses the horizon and tunnels out, it escapes to infinity and appear
as Hawking radiation. Meanwhile, when a particle with negative energy
tunnels inwards it is absorbed by the black hole and as a result the
mass of the black hole decreases. Therefore, the essence of the quantum
tunneling argument for Hawking radiation is the calculation of the
imaginary part of the action. If we consider the action $\mathcal{I}\left(r,t\right)$
and make an expansion in powers of $\hbar$ we obtain

\begin{eqnarray}
\mathcal{I}\left(r,t\right) & = & \mathcal{I}_{0}\left(r,t\right)+\hbar\mathcal{I}_{1}\left(r,t\right)+\hbar^{2}\mathcal{I}_{2}\left(r,t\right)+...\\
 & = & \mathcal{I}_{0}\left(r,t\right)+\sum_{i}\hbar^{i}\mathcal{I}_{i}\left(r,t\right),\end{eqnarray}

where $\mathcal{I}_{0}$ gives the semiclassical value and the terms
from $O\left(\hbar\right)$ onwards are treated as quantum corrections.
The work of Banerjee and Majhi \cite{expansion} shown that the correction
terms $\mathcal{I}_{i}$ are proportional to the semiclassical contribution
$\mathcal{I}_{0}$. Since $\mathcal{I}_{0}$ has the dimension of
$\hbar$, the proportionality constants should have the dimension
of inverse of $\hbar$. In natural units $\left(G=c=k_{B}=1\right)$,
the Planck constant is of the order of square of the Planck Mass.
Therfore, from dimensional analysis the proportionality constants
have the dimension of $M^{-2i}$, where $M$ is the mass of black
hole, and the series expansion becomes

\begin{eqnarray}
\mathcal{I}\left(r,t\right) & = & \mathcal{I}_{0}\left(r,t\right)+\sum_{i}\beta_{i}\frac{\hbar^{i}}{M^{2i}}\mathcal{I}_{0}\left(r,t\right)\\
 &  & \mathcal{I}_{0}\left(r,t\right)\left(1+\sum_{i}\beta_{i}\frac{\hbar^{i}}{M^{2i}}\right),\end{eqnarray}
where $\beta_{i}$'s are dimensionless constant parameters. If the
black hole has other macroscopic parameters such as angular momentum
or electric charge, one can express this expansion in terms of those
parameters, as done in  \cite{otros} and \cite{akbar}. In this
work, the dimensional analysis suggest that the constants of proportionality
for charged rotating black holes have the dimensions of $\left(r_{H}^{2}-2Q^{2}e^{-2\phi_{0}}\right)^{i}$,
so we will use an expansion in terms of the horizon radius and the
electric charge as

\begin{eqnarray}
\mathcal{I}\left(r,t\right) & = & \mathcal{I}_{0}\left(r,t\right)\left(1+\sum_{i}\beta_{i}\frac{\hbar^{i}}{\left(r_{H}^{2}-2Q^{2}e^{-2\phi_{0}}\right)^{i}}\right).\label{eq:actionexpansion}\end{eqnarray}

Using this expansion we will calculate the quantum corrections to
the entropy of the GMGHS black hole.

\section{Entropy as an Exact Differential}

In order to perform the quantum corrections to the entropy of the
black hole we will follow the analysis of \cite{otros,akbar}. The first
law of thermodynamics for charged black holes is 

\begin{equation}
dM=TdS+\Phi dQ,\end{equation}
where the parameters $M$ and $Q$ are the mass and charge of the
black hole, while $T,S$ and $\Phi$ are the temperature, entropy
and electrostatic potential, respectively. This equation can be re-written
as

\begin{equation}
dS\left(M,Q\right)=\frac{1}{T}dM-\frac{\Phi}{T}dQ,\end{equation}
from which one can infer that in order for $dS$ to be an exact differential,
the thermodynamical quantities must satisfy

\begin{eqnarray}
\frac{\partial}{\partial Q}\left(\frac{1}{T}\right) & = & \frac{\partial}{\partial M}\left(-\frac{\Phi}{T}\right).\label{eq:cond1}\end{eqnarray}

If $dS$ is an exact differential, we can write the entropy $S(M,J,Q)$
in the integral form

\begin{eqnarray}
S\left(M,Q\right) & = & \int\frac{1}{T}dM-\int\frac{\Phi}{T}dQ-\int\left(\frac{\partial}{\partial Q}\left(\int\frac{1}{T}dM\right)\right)dQ.\label{eq:integralentropy}\end{eqnarray}

\section{Standard Entropy of the GMGHS Black Hole}

The low energy effective action of the heterotic string theory in
four dimensions is given by

\begin{equation}
\mathcal{A}=\int d^{4}x\sqrt{-g}e^{-\phi}\left(-R+\frac{1}{12}H_{\mu\nu\rho}H^{\mu\nu\rho}-G^{\mu\nu}\partial_{\mu}\phi\partial_{\nu}\phi+\frac{1}{8}F_{\mu\nu}F^{\mu\nu}\right),\end{equation}

where $R$ is the Ricci scalar, $G_{\mu\nu}$ is the metric that arises
naturally in the $\sigma$ model, 

\begin{equation}
F_{\mu\nu}=\partial_{\mu}A_{\nu}-\partial_{\nu}A_{\mu}\end{equation}
is the Maxwell field associated with a $U\left(1\right)$ subgroup
of $E_{8}\times E_{8}$, $\phi$ is the dilaton field and

\begin{equation}
H_{\mu\nu\rho}=\partial_{\mu}B_{\nu\rho}+\partial_{\nu}B_{\rho\mu}+\partial_{\rho}B_{\mu\nu}-\left[\Omega_{3}\left(A\right)\right]_{\mu\nu\rho},\end{equation}
where $B_{\mu\nu}$ is the antisymmetric tensor gauge field and

\begin{equation}
\left[\Omega_{3}\left(A\right)\right]_{\mu\nu\rho}=\frac{1}{4}\left(A_{\mu}F_{\nu\rho}+A_{\nu}F_{\rho\mu}+A_{\rho}F_{\mu\nu}\right)\end{equation}
is the gauge Chern-Simons term. Considering $H_{\mu\nu\rho}=0$ and
working in the conformal Einstein frame, the action becomes

\begin{equation}
\mathcal{A}=\int d^{4}x\sqrt{-g}\left(-R+2\left(\nabla\phi\right)^{2}+e^{-2\phi}F^{2}\right),\end{equation}
where the Einstein frame metric $g_{\mu\nu}$ is related to $G_{\mu\nu}$
through the dilaton,

\begin{equation}
g_{\mu\nu}=e^{-\phi}G_{\mu\nu}.\end{equation}

The charged black hole solution, known as the Gibbons-Maeda- Garfinkle-Horowitz-Strominger
(GMGHS) solution, is given by \cite{gmghs,shaoWEn}

\begin{equation}
ds^{2}=-\left(1-\frac{2M}{r}\right)dt^{2}+\left(1-\frac{2M}{r}\right)^{-1}dr^{2}+r^{2}\left(1-\frac{Q^{2}e^{-2\phi_{0}}}{Mr}\right)d\Omega^{2}\end{equation}
where

\begin{equation}
e^{-2\phi}=e^{-2\phi_{0}}\left(1-\frac{Q^{2}e^{-2\phi_{0}}}{Mr}\right)\end{equation}

\begin{equation}
F=Q\sin\theta d\theta\wedge d\varphi\end{equation}
and $\phi_{0}$ is the asymptotic constant value of $\phi$ at $r\rightarrow\infty$.
Note that this metric become Schwarzschild\textasciiacute{}s solution
if $Q=0$. The GMGHS solution has a spherical event horizon at

\begin{equation}
r_{H}=2M\label{eq:horizon}\end{equation}

and its area is given by

\begin{equation}
A=\int\sqrt{g_{\theta\theta}g_{\varphi\varphi}}d\theta\varphi=4\pi\left(r_{H}^{2}-2Q^{2}e^{-2\phi_{0}}\right).\label{eq:area}\end{equation}

Equation (\ref{eq:area}) tell us that the area of the horizon goes
to zero if

\begin{equation}
r_{H}^{2}=2Q^{2}e^{-2\phi_{0}},\end{equation}
i.e. the GMGHS solution becomes a naked singularity if

\begin{equation}
M^{2}\leq\frac{1}{2}Q^{2}e^{-2\phi_{0}}.\end{equation}

The Hawking temperature is

\begin{equation}
T_{H}=\frac{\kappa\hbar}{2\pi}=\frac{\hbar}{8\pi M},\end{equation}

which is independent of charge. Finally, the electric potential computed
on the horizon of the black hole is 

\begin{equation}
\Phi=\frac{Q}{r_{H}}e^{-2\phi_{0}}.\end{equation}
One can easily check that thermodynamical quantities for the GMGHS
black hole satisfy condition (\ref{eq:cond1}), making $dS$ an exact
differential. Thus, the integral form of the entropy (\ref{eq:integralentropy})
gives 

\begin{equation}
S_{0}\left(M,Q\right)=\int\frac{1}{T_{H}}dM-\int\frac{\Phi}{T_{H}}dQ\end{equation}

\begin{equation}
S_{0}\left(M,Q\right)=\frac{4\pi}{\hbar}\left[M^{2}-\frac{e^{-2\phi_{0}}}{2}Q^{2}\right]\end{equation}
that corresponds to the standard black hole entropy 

\begin{equation}
S_{0}\left(M,Q\right)=\frac{A}{4\hbar}=\frac{\pi\left(r_{H}^{2}-2Q^{2}e^{-2\phi_{0}}\right)}{\hbar}.\label{eq:standardentropy}\end{equation}

\section{Quantum Correction of the Entropy}

When considering the expansion for the action (\ref{eq:actionexpansion}),
it affects the Hawking temperature by introducing some correction
terms \cite{otros,akbar,expansion}. Therefore the temperature is now given
by

\begin{eqnarray}
T & = & T_{H}\left(1+\sum_{i}\beta_{i}\frac{\hbar^{i}}{\left(r_{H}^{2}-2Q^{2}e^{-2\phi_{0}}\right)^{i}}\right)^{-1},\label{eq:newtemperature}\end{eqnarray}
where $T_{H}$ is the standard Hawking temperature and the terms with
$\beta_{i}$ are quantum corrections to the temperature. It is not
difficult to verify that the conditions to make $dS$ an exact differential
are satisfied when considering this new form for the temperature.
Therefore, the entropy with correction terms is given by

\begin{equation}
S\left(M,Q\right)=\int\frac{1}{T}dM=\int\frac{1}{T_{H}}\left(1+\sum_{i}\beta_{i}\frac{\hbar^{i}}{\left(r_{H}^{2}-2Q^{2}e^{-2\phi_{0}}\right)^{i}}\right)dM\end{equation}
or 

\begin{equation}
S\left(M,Q\right)=\int\frac{1}{T_{H}}dM+\int\frac{\beta_{1}}{T_{H}}\frac{\hbar}{\left(r_{H}^{2}-2Q^{2}e^{-2\phi_{0}}\right)}dM+\int\frac{\beta_{2}}{T_{H}}\frac{\hbar^{2}}{\left(r_{H}^{2}-2Q^{2}e^{-2\phi_{0}}\right)^{2}}dM+...\end{equation}
This equation can be written as 

\begin{equation}
S\left(M,Q\right)=S_{0}+S_{1}+S_{2}+....,\end{equation}
where $S_{0}$ is the standard entropy given by equation (\ref{eq:standardentropy})
and $S_{1},S_{2},...$ are quantum corrections. The first of these
terms is

\begin{eqnarray}
S_{1} & = & \beta_{1}\hbar\int\frac{1}{T_{H}\left(r_{H}^{2}-2Q^{2}e^{-2\phi_{0}}\right)}dM.\end{eqnarray}
Solving the integral, we obtain

\begin{equation}
S_{1}=\pi\beta_{1}\ln\left|r_{H}^{2}-2Q^{2}e^{-2\phi_{0}}\right|.\end{equation}
The following terms can be written, in general, as

\begin{equation}
S_{j}=\beta_{j}\hbar^{j}\int\frac{1}{T_{H}\left(r_{H}^{2}-2Q^{2}e^{-2\phi_{0}}\right)^{j}}dM\end{equation}
By calculating the integral, we obtain

\begin{equation}
S_{j}=\frac{\pi\beta_{j}\hbar^{j-1}}{1-j}\left(r_{H}^{2}-2Q^{2}e^{-2\phi_{0}}\right)^{1-j}\end{equation}
for $j>1$. Therefore, the entropy with quantum corrections is given
by

\begin{eqnarray}
S\left(M,Q\right) & = & \frac{\pi\left(r_{H}^{2}-2Q^{2}e^{-2\phi_{0}}\right)}{\hbar}+\pi\beta_{1}\ln\left|r_{H}^{2}-2Q^{2}e^{-2\phi_{0}}\right|\nonumber \\
 &  & +\sum_{j>1}\frac{\pi\beta_{j}\hbar^{j-1}}{1-j}\left(r_{H}^{2}-2Q^{2}e^{-2\phi_{0}}\right)^{1-j}.\end{eqnarray}

Using equation (\ref{eq:area}), and doing a re-definition of the
$\beta_{i}$, we can write the entropy in terms of the area of the
horizon as

\begin{eqnarray}
S\left(M,Q\right) & = & \frac{A}{4\hbar}+\pi\beta_{1}\ln\left|A\right|+\sum_{j>1}\frac{\pi\beta_{j}\hbar^{j-1}}{1-j}\left(\frac{A}{4\pi}\right)^{1-j}.\end{eqnarray}

The first term in this expansion is the usual semiclassical entropy
while the second term is the logarithmic correction found earlier
for some geometries and using different methods \cite{kumar,larranaga}
. The value of the coefficients $\beta_{i}$ can be evalated using
other approaches, such as the entanglement entropy calculation. Finally
note that the third term in the expansion is an inverse of area term
similar to the one obtained by S. K. Modak \cite{kumar} for the rotating
BTZ black hole, for the charged BTZ black hole \cite{larranaga} and
also in the general cases studied in \cite{otros} and \cite{akbar}.

\section{Conclusion}

As is well known, the Hawking evaporation process can be understood
as a consequence of quantum tunneling in which some particles cross
the event horizon. The positive energy particles tunnel out of the
event horizon, whereas, the negative energy particles tunnel in, resulting
in black hole evaporation. Using this analysis we have studied the
quantum corrections to the entropy of the GMGHS black hole of heterotic
string theory. With the help of the conditions for exactness of differential
of entropy we obtain a power series for entropy. The first term is
the semiclassical value, while the leading correction term is logarithmic
as has been found using other methods\cite{kumar,larranaga}. The
other terms involve ascending powers of the inverse of the area. This
analysis shows that the quantum corrections to entropy obtained in the literature
 \cite{otros,akbar}, also hold for the black hole of
string theory studied here. 

\emph{Acknowledgements}. This work was supported by the Universidad
Nacional de Colombia. Project Code 2010100.

\end{document}